\documentstyle[preprint,eqsecnum,aps]{revtex} 
\begin{document}
\preprint{ PUPT-1753 IASSNS-HEP-97-117}
\title{An Action for Black Hole Membranes}
\author{Maulik K. Parikh}
\address{Joseph Henry Laboratories, Princeton University, Princeton, New 
Jersey 08544, USA}
\author{Frank Wilczek}
\address{School of Natural Sciences, Institute for Advanced Study, Princeton, 
New Jersey 08540, USA}
\date{\today}
\maketitle
\begin{abstract}
The membrane paradigm is the remarkable view that, to an external observer, a 
black hole appears to behave exactly like a
dynamical fluid membrane, obeying such pre-relativistic equations as Ohm's
law and the Navier-Stokes equation. 
It has traditionally been derived by manipulating the equations of motion.
Here we provide an action formulation of
this picture, clarifying what underlies the paradigm, and simplifying the
derivations. Within this framework, we derive previous membrane
results, and extend them to dyonic black hole solutions.
We discuss how it is that an action can produce dissipative equations. 
Using a Euclidean path integral,
we show that familiar semi-classical thermodynamic properties of black
holes also emerge from the membrane action. Finally, in a Hamiltonian
description, we establish the validity of a minimum entropy production
principle for black holes.

\end{abstract}

\def \lm {\lambda}
\def \gm {\gamma}
\def \d {\delta}
\def \pl {\partial}
\def \cc {\overline}
\def \l2 {\lambda _2}
\def \f {\frac}
\def \del {\nabla}
\def \rg {\sqrt {-g} \,}
\def \rh {\sqrt {-h} \,}
\def \mn {\mu \nu}
\def \lf {\left (}
\def \rt {\right )}

\section{INTRODUCTION}         
The event horizon of a black hole is a peculiar object: it is a mathematically
defined, locally undetectable boundary, a surface-of-no-return 
inside which light cones tip over and
``time'' becomes spatial  \cite{he}.
Otherwise natural descriptions of physics often have trouble 
accommodating the horizon; as the most primitive example, the 
familiar Schwarzschild metric has a co-ordinate singularity there. 
Theories of fields that extend to the horizon face the additional challenge of
having to define boundary conditions on a surface that is infinitely
red-shifted, has a singular Jacobian, and possesses a normal vector which is
also tangential. These considerations might induce one to believe that
black hole horizons are fundamentally different from other physical entities.

On the other hand, further work has established a 
great variety of analogies between the
horizon and more familiar, pre-relativistic bodies. In addition to
the famous four laws of black hole thermodynamics \cite{ch,chr,bek,swh},
which are global statements, there is also a precise local mechanical and
electrodynamic correspondence. In effect, it has been
shown \cite{hr,td78,zna,td79,td82} that an observer who remains
outside a black hole perceives the
horizon to behave according to equations that describe a fluid bubble with
electrical conductivity as well as shear and bulk viscosities.
Moreover, it is possible to define a set of local surface densities, such as
charge or energy-momentum, which
inhabit the bubble surface and which obey conservation laws.
Quite remarkably, a general-relativistically
{\em exact} calculation then leads, for arbitrary non-equilibrium
black holes, to equations for the horizon which can be
precisely identified with Ohm's law,
the Joule heating law, and the Navier-Stokes equation.

These relations were originally derived for the mathematical, or true, event
horizon. For astrophysical applications it became more convenient to consider
instead a ``stretched horizon,'' a 2+1-dimensional time-like surface located 
slightly outside the true horizon.
Because it has a non-singular induced metric, the stretched horizon
provides a more tractable boundary on which to anchor external fields; 
outside a complicated boundary layer, the equations governing the 
stretched horizon are to excellent approximation \cite{tmac,pt}
the same as those for the true horizon. This view of
a black hole as a dynamical time-like surface, or membrane, has been called
the membrane paradigm \cite{mp}.

Most of the mentioned results have been derived through general-relativistic
calculations based on various intuitive physical arguments. In this paper,
we show that the gravitational and electromagnetic descriptions of the
membrane can be derived systematically, directly, and more simply 
from the Einstein-Hilbert or Maxwell actions. Aside from the appeal 
inherent in a least action
principle, an action formulation is a unifying framework which is easily
generalizable and has the advantage of providing a bridge to thermodynamics
and quantum mechanics (see \cite{carlip} for related work).
In a follow-up paper, we exploit these advantages
to evaluate some effects of back-reaction of spacetime geometry 
on Hawking radiation \cite{tunnel}.

The key idea in what follows is that, since
(classically) nothing can emerge from a black hole, 
an observer who remains outside a black hole cannot be affected by the 
dynamics inside the hole. Hence the equations of motion ought to
follow from varying an action restricted to the external
universe. However, the boundary term in the derivation of the
Euler-Lagrange equations does not in general vanish on the stretched
horizon as it does at the boundary of spacetime. In order to obtain the correct
equations of motion, we must add to the external action a surface term
that cancels this residual boundary term. The membrane picture emerges
in interpreting the added surface term as electromagnetic and 
gravitational sources residing on the stretched horizon.

In the rest of this paper, we examine individually the boundary
terms for the electromagnetic, gravitational, and axidilaton cases.
We also discuss dissipation and provide thermodynamic and Hamiltonian
descriptions. We use lowercase indices for four-dimensional tensor indices 
and uppercase indices for
the two-dimensional tensors that occupy space-like sections of the horizon.
We use geometrized units ($G \equiv c \equiv 1$), and a 
spacetime metric with positive signature $\lf -+++ \rt$. Our sign
conventions are those of MTW \cite{mtw}, with the 
exception of the extrinsic curvature
which we define to have positive trace for a convex surface. 

\section{HORIZON PRELIMINARIES}
In this section, we fix our conventions, first in words, then in equations. 
Through every point on the true horizon there exists a unique null
generator $l^a$ which we may parameterize by some regular time co-ordinate
whose normalization we fix to equal that of time-at-infinity.
Next, we choose a time-like surface just outside the true horizon. This is
the stretched horizon, $\cal{H}$, whose location we parameterize by 
$\alpha \ll 1$ so that
$\alpha \to 0$ is the limit in which the stretched horizon coincides
with the true horizon. We will always take this limit at the end of any
computation. Since many of the useful intermediate quantities will 
diverge as inverse powers of $\alpha$, we renormalize them by the 
appropriate power of $\alpha$. In
that sense, $\alpha$ plays the role of a regulator.

For our purposes, the principal reason
for preferring the stretched horizon over the true horizon is that the
metric on a time-like - rather than null - surface is non-degenerate, 
permitting one to write down a conventional action. Generically (in the
absence of horizon caustics), a one-to-one correspondence
between points on the true and stretched horizons is always possible
via, for example, ingoing 
null rays that pierce both surfaces (see \cite{pt} for details).

We can take the stretched horizon to be the world-tube of a family of
time-like observers who hover just outside 
the true horizon. These nearly light-like 
``fiducial'' observers are pathological in that
they suffer an enormous proper acceleration and measure quantities that
diverge as $\alpha \to 0$. However, although we 
take the mathematical limit in
which the true and stretched horizons conflate, for physical purposes 
the proper distance of 
the stretched horizon from the true horizon need only be smaller 
than the length scale involved in a given measurement.
In that respect, the stretched horizon, although a
surrogate for the true horizon, is actually more fundamental
than the true horizon, since measurements at the stretched horizon constitute
real measurements that an external observer could make and report, whereas 
accessing any quantity measured at the true horizon would entail the observer's
inability to report back his or her results.

We take our fiducial observers to have world lines $U^a$, parameterized by
their proper time, $\tau$. The stretched horizon also possesses a space-like 
unit normal $n^a$ which for
consistency we shall always take to be outward-pointing. Moreover, we choose
the normal vector congruence on the stretched horizon to emanate outwards
along geodesics. We define $\alpha$ by requiring that $\alpha U^a 
\to l^a$ and 
$\alpha n^a \to l^a$; hence $\alpha U^a$ and $\alpha n^a$ are equal in
the true horizon limit. This is nothing more than the statement that the
null generator $l^a$ is both normal and tangential to the true horizon, which
is the defining property of null surfaces.
Ultimately though, it will be this property that will be 
responsible for the dissipative behavior of the horizons.
The 3-metric, $h_{ab}$, on $\cal{H}$ can 
be written as a 4-dimensional tensor
in terms of the spacetime metric and the normal vector, so that
$h^a_b$ projects from the spacetime tangent space to the 3-tangent space. 
Similarly, we can
define the 2-metric, $\gm _{AB}$, of the spacelike section of $\cal{H}$ 
to which
$U^a$ is normal, in terms of the stretched horizon 3-metric and $U^a$, thus
making a 2+1+1 split of spacetime. We denote the 4-covariant derivative
by $\del_a$, the 3-covariant derivative by $_{|a}$, and the 2-covariant
derivative by $_{\|A}$. For a vector in the stretched horizon, the
covariant derivatives are related by $h^c_d \del _c w^a = w^a_{|d} -
K^c_d w_c n^a$ where $K^a_b \equiv h^c_b \del_c n^a$ is the stretched
horizon's extrinsic curvature, or second fundamental form.
In summary,
\begin{equation}
l^2 = 0
\end{equation}
\begin{equation}
U^a = \lf \f{d}{d \tau} \rt ^a \; , \; \; U^2 = -1 \; , \; \; 
\lim_{\alpha\to\infty} \alpha U^a = l^a
\end{equation}
\begin{equation}
n^2 = + 1 \; , \; \; a^c = n^a \del _a n^c = 0 \; , \; \; 
\lim_{\alpha\to\infty} \alpha n^a = l^a
\end{equation}
\begin{equation}
h^a_b = g^a_b - n^a n_b \; , \; \; \gm^a_b = h^a_b + U^a U_b 
= g^a_b - n^a n_b + U^a U_b
\end{equation}
\begin{equation}
K^a_b \equiv h^c_b \del_c n^a \; , \; \; K_{ab} = K_{ba} \; , \; \; 
K_{ab} n^b = 0
\end{equation}
\begin{equation}
w^c \epsilon {\cal{H}} \Rightarrow h^c_d \del _c w^a = w^a_{|d} -
K^c_d w_c n^a \Rightarrow \del_c w^c = w^c _{|c} + w^c a_c = w^c _{|c}
\; . \label{div}
\end{equation}
The last expression relates the covariant divergence associated with $g_{ab}$
to the covariant divergence associated with $h_{ab}$.

For example, the Reissner-Nordstr\"{o}m solution has 
\begin{equation}
ds^2 = - \lf 1 - \f{2M}{r} + \f{Q^2}{r^2} \rt dt^2 + 
{\lf 1 - \f{2M}{r} + \f{Q^2}{r^2} \rt}^{-1} dr^2 + r^2 d \Omega ^2 \; ,
\end{equation}
so that a stretched horizon at constant $r$ would have
\begin{equation}
\alpha = {\lf 1 - \f{2M}{r} + \f{Q^2}{r^2} \rt}^{\f{1}{2}} \; ,
\end{equation}
\begin{equation}
U_a = - \alpha \lf d t \rt _a \; ,
\end{equation}
and
\begin{equation}
n_a = + \alpha ^{-1} \lf d r \rt _a \; .
\end{equation}

\section{THE ACTIONS}
To find the complete equations of motion by extremizing an action,
it is not sufficient to set the bulk
variation of the action to zero: one also needs to use the boundary 
conditions. Here we take our Dirichlet boundary conditions 
to be $\d \phi = 0$ at the singularity and at the boundary of spacetime,
where $\phi$ stands for any field. 

Now since the fields inside a black hole cannot have any 
classical relevance for an external observer,
the physics must follow from varying the part of the action restricted to the
spacetime outside the black hole. However, this external action is not
stationary on its own, because boundary conditions are fixed only at the
singularity and at infinity, but not at the stretched horizon. Consequently, we
re-write the total action as

\begin{equation}
S_{world} = \lf S_{out} + S_{surf} \rt + \lf S_{in} -S_{surf} \rt \; ,
\label{split}
\end{equation}
where now $\d S_{out} + \d S_{surf} \equiv 0$, which implies also that $\d 
S_{in} - \d S_{surf} = 0$. The total action has been broken down into two 
parts,
both of which are stationary on their own, and which do not require any
new boundary conditions.

The surface term, $S_{surf}$, corresponds to sources, such as surface 
electric charges and currents for the Maxwell action, or surface stress
tensors for the Einstein-Hilbert action. The sources are fictitious: an
observer who falls through the stretched horizon will not
find any surface sources and, in fact, will not find any stretched horizon.
Furthermore, the field configurations inside the black hole will be 
measured by
this observer to be entirely different from those posited by the membrane
paradigm. On the other hand, for an external fiducial observer the
source terms are
a very useful artifice; their presence is consistent with all external
fields.
This situation is directly analogous to the method
of image charges in electrostatics, in which a fictitious charge
distribution is added to the system to implement, say, conducting boundary
conditions.
By virtue of the uniqueness of solutions to Poisson's equation with
conducting boundary conditions, the electric
potential on one - and only one - side of the boundary is guaranteed to
be the correct
potential. An observer who remains on that side of the boundary has no way
of telling through the fields alone whether they arise through the fictitious
image charges or through actual surface charges. The illusion is exposed only
to the observer who crosses the boundary to find that not only are there
no charges, but the potential on the other side of the boundary is
quite different from what it would have been had the image charges been real.

In the rest of this section, we shall implement eqn. \ref{split} concretely in 
important special cases.

\subsection{The Electromagnetic Membrane}
The external Maxwell action is
\begin{equation}
S_{out}[A_a] = \int d^4 x \rg \lf - \f {1}{16 \pi} F^2 + J \cdot A \rt \; ,
\end{equation}
where $F$ is the electromagnetic field strength.
Under variation, we obtain the inhomogeneous Maxwell equations,
\begin{equation}
\del_b F^{ab} = 4 \pi J^a \; ,
\end{equation}
as well as the boundary term
\begin{equation}
\f {1} {4 \pi} \int d^3 x \rh F^{ab} n_a \d A_b \; ,
\end{equation}
where $h$ is the determinant of the induced metric, 
and $n^a$ is the outward-pointing space-like unit normal to the
stretched horizon. We need to cancel this term. Adding the surface term
\begin{equation}
S_{surf}[A_a] = + \int d^3 x \rh j_s \cdot A \; ,		\label{charge}
\end{equation}
we see that we must have
\begin{equation}
j_s^a = + \f {1} {4 \pi} F^{a b} n_b \; .	\label{j}
\end{equation}
The surface 4-current, $j_s^a$,
has a simple physical interpretation. We see 
that its time-component is a surface charge, $\sigma$, 
that terminates the normal component of the electric 
field just outside the membrane, while the spatial
components, $\vec{j_s}$, form a surface current 
that terminates the tangential component
of the external magnetic field:
\begin{equation}
E_{\perp} = -U_a F^{ab} n_b = 4 \pi \sigma
\end{equation}
\begin{equation}
\vec{B}_{\|}^A = \epsilon^A_B \gm^B_a F^{ab} n_b = 4 \pi \lf \vec{j}_s 
\times \hat{n} \rt ^A	\; . \label{B}
\end{equation}
It is characteristic of the membrane paradigm that $\sigma$ and $\vec{j}_s$ are
{\em local} densities, so that the total charge on the black hole is the
surface integral
of $\sigma$ over the membrane, taken at some constant universal time.
This is in contrast to the total
charge of a Reissner-Nordstr\"{o}m black hole, which is a 
global characteristic that can be defined by an integral at spatial infinity.

{}From Maxwell's equations and eqn. \ref{j}, we obtain a continuity 
equation for
the membrane 4-current which, for a stationary hole, takes the form
\begin{equation}
\f {\pl \sigma}{\pl \tau} + \vec{\del}_2 \cdot \vec{j}_s  = - J^n \; ,
\end{equation}
where $\vec{\del}_2 \cdot \vec{j}_s \equiv {\lf \gm^A_a j_s^a \rt}_{\|A}$ 
is the two-dimensional divergence of 
the membrane surface current, and $- J^n = -J^a n_a$ is the amount of charge
that falls into the hole per unit area per unit proper time, $\tau$.
Physically, this equation expresses local charge conservation in that 
any charge that falls into the black hole can be regarded as remaining on 
the membrane: the membrane is impermeable to charge.

The equations we have so far are sufficient to determine the fields outside
the horizon, given initial conditions outside the horizon.
A plausible requirement for initial conditions at the horizon is that the
fields measured by freely-falling observers (FFO's) at the stretched horizon 
be finite. 
There being no curvature singularity at the horizon, inertial observers
who fall through the horizon should detect nothing out of the ordinary. In
contrast, the fiducial observers (FIDO's) who make measurements at the membrane
are infinitely accelerated. Their measurements, subject to infinite
Lorentz boosts, are singular. For the electromagnetic fields we
have, with $\gm$ the Lorentz boost and using orthonormal co-ordinates,
\begin{eqnarray}
E_{\theta}^{FIDO} \approx  \gm \lf E_{\theta}^{FFO} - B_ 
{\phi}^{FFO} \rt \; , \; \; &
B_{\phi}^{FIDO} \approx  \gm \lf B_{\phi}^{FFO} - E_
{\theta}^{FFO} \rt \; , \\
B_{\theta}^{FIDO} \approx  \gm \lf B_{\theta}^{FFO} - E_
{\phi}^{FFO} \rt \; , \; \; &
E_{\phi}^{FIDO} \approx  \gm \lf E_{\phi}^{FFO} - B_
{\theta}^{FFO} \rt \; ,
\end{eqnarray}
or, more compactly,
\begin{equation}
\vec{E}^{FIDO}_{\|} = \hat{n} \times \vec{B}^{FIDO}_{\|} \; .	\label{reg}
\end{equation}
That is, the regularity condition states that all radiation in the 
normal direction is ingoing; a black hole acts as a perfect absorber.
Combining the regularity condition with eqn. \ref{B} and dropping the FIDO
label, we arrive at
\begin{equation}
\vec{E}_{\|} = 4 \pi \vec{j}_s	\; .
\end{equation}
That is, black holes obeys Ohm's law with a surface resistivity
of $\rho = 4 \pi \approx 377 \, \Omega$. Furthermore, the Poynting flux is
\begin{equation}
\vec{S} = \f {1} {4 \pi} \lf \vec{E} \times \vec{B} \rt = - j_s^2 \rho \, 
\hat{n} \; .
\end{equation}
We can integrate this over the black hole horizon at some fixed
time. However, for a generic stretched horizon, we cannot
time-slice using fiducial time as different fiducial 
observers have clocks that do not necessarily remain synchronized. Consequently
we must use some other time for slicing purposes, such as the time at
infinity, and then include in the integrand a (potentially
position-dependent) factor to convert the locally measured energy flux
to one at infinity. With a clever choice of the stretched horizon,
however, it is possible to arrange that all fiducial observers have
synchronized clocks. In this case, two powers of $\alpha$, which is
now the lapse, are included in the integrand. 
Then, for some given universal time, $t$, the power radiated into
the black hole, which is also the rate of increase of the black hole's
irreducible mass, is given by
\begin{equation}
\f {d M_{irr}} {d t} = - \int \alpha^2 \vec{S} \cdot d \vec{A} = 
+ \int \alpha^2 j_s^2 \rho \, dA \; .
\end{equation}
That is, black holes obey the Joule heating law, the same law that also
describes the dissipation of an ordinary ohmic resistor.

\subsection{The Gravitational Membrane}
We turn now to gravity. The external Einstein-Hilbert action is
\begin{equation}
S_{out}[g^{ab}] = \f {1} {16 \pi} \int d^4 x \rg R + \f {1} {8 \pi} \oint
d^3 x \sqrt {\pm h} \, K + S_{matter} \; ,
\end{equation}
where $R$ is the Ricci scalar, $K$ is the trace of the extrinsic curvature, 
and where for convenience we have chosen the
field variable to be the inverse metric $g^{ab}$.
The surface integral of $K$ is only over the outer boundary of spacetime, 
and not over the stretched horizon. It is
required in order to obtain the Einstein equations because the Ricci scalar
contains second order derivatives of $g_{ab}$. When this action is varied,
the bulk terms give the Einstein equations,
\begin{equation}
R_{ab} - \f {1}{2} g_{ab} R = 8 \pi T_{ab} \; . \label{ein}
\end{equation}
We are interested however in the interior boundary term. This comes from the
variation of the Ricci tensor. We note that
\begin{equation}
g^{ab} \d R_{ab} = \del ^a \lf \del ^b \lf \d g_{ab} \rt - g^{cd} \del _a \lf
 \d g_{cd} \rt \rt \; ,
\end{equation}
where $\d g_{ab} = - g_{ac} g_{bd} \d g^{cd}$. Gauss' theorem now gives
\begin{equation}
\int d^4 x \rg \lf g^{ab} \d R_{ab} \rt = - \int d^3 x \rh n^a h^{bc} \lf 
\del_c \lf \d g_{ab} \rt - \del_a \lf \d g_{bc} \rt \rt \; ,
\end{equation}
where the minus sign arises from choosing $n^a$ to be 
outward-pointing. Applying the Leibniz rule, we can rewrite this as
\begin{eqnarray}
\int d^4 x \rg \lf g^{ab} \d R_{ab} \rt & = & \int d^3 x \rh h^{bc} \lf 
\del _a \lf n^a \d g_{bc} \rt 
{} - \d g_{bc} \del _a \lf n^a \rt \right. \nonumber \\
& & \indent \indent \; \; \left. 
{} - \del _c \lf n^a \d g_{ab} \rt
+ \d g_{ab} \del _c \lf n^a \rt \rt \; .	\label{long}
\end{eqnarray}

Now, in the limit that the stretched horizon approaches the null horizon, the
first and third terms on the right-hand side vanish:
\begin{equation}
\int d^3 x  \rh h^{bc} \lf \del _a \lf n^a \d g_{bc} \rt 
{} - \del _c \lf n^a \d g_{ab} \rt \rt = 0 \; .
\label{zero}
\end{equation}
A proof of this identity is given in the appendix. With $K^{ba} =
h^{bc} \del_c n^a$, the variation of the external action is 
\begin{equation}
\d S_{out}[g^{ab}] = \f {1}{16 \pi} \int d^3 x \rh \lf K h_{ab} - K_{ab} \rt 
\d g^{ab} \; .	\label{var}
\end{equation}
Since the expression in parentheses contains only stretched horizon tensors,
the normal vectors in the variation $\d g^{ab} = \d h^{ab} + 
\d n^a n^b + n^a \d n^b$ contribute nothing.
As in the electromagnetic case, we add a surface source term to the action
to cancel this residual boundary term. The variation of the required term 
can therefore be written as
\begin{equation}
\d S_{surf}[h^{ab}] = - \f {1} {2} \int d^3 x \rh t_{s \, ab} \d
h^{ab} \; . \label{cancel}
\end{equation}
We shall see later that this variation is integrable i.e. an
action with this variation exists. Comparison with equation \ref{var}
yields the membrane stress tensor:
\begin{equation}
t_s^{ab} = + \f {1} {8 \pi} \lf K h^{ab} - K^{ab} \rt \; . \label{surf}
\end{equation}
Now just as a surface charge produces a discontinuity in the normal component
of the electric field, a surface stress term creates a discontinuity in the
extrinsic curvature. The relation between the discontinuity and the source
term is given by the Israel junction condition \cite{mtw},
\begin{equation}
t_s^{ab} = \f {1} {8 \pi} \lf [K] h^{ab} - [K]^{ab} \rt	\; ,
\end{equation}
where $[K] = K_+ - K_-$ is the difference in the extrinsic curvature of the
stretched horizon between its embedding in the external universe and its
embedding in the spacetime internal to the black hole.
Comparing this with our result for the membrane stress tensor, eqn. 
\ref{surf}, we see that
\begin{equation}
K_- ^{ab} = 0 \; ,
\end{equation}
so that the interior of the stretched horizon molds itself into flat space.
The Einstein equations, eqn. \ref{ein}, can be rewritten
via the contracted Gauss-Codazzi equations \cite{mtw} as
\begin{equation}
t^{ab}_{s \; \: |b} = - h^a_c T^{cd} n_d \; .	\label{gcod}
\end{equation}

Eqns. \ref{surf} and \ref{gcod}
taken together imply that the stretched horizon can be thought of as a fluid
membrane, obeying the Navier-Stokes equation. To see this, recall that
as we send $\alpha$ to zero, both $\alpha U^a$ and $\alpha n^a$ approach $l^a$,
the null generator at the corresponding point on the true horizon. 
Hence, in this limit we can equate $\alpha U^a$ and $\alpha n^a$, permitting us
to write the relevant components of $K^a_b$, in terms of the surface
gravity, $g$, and the extrinsic curvature, $k^A_B$,
of a spacelike 2-section of the stretched horizon:
\begin{equation}
U^c \del _c n^a \to \alpha ^{-2} l^c \del _c l^a 
\equiv \alpha ^{-2} g_H l^a
\Rightarrow K_a^b U^a U_b = - g = - \alpha ^{-1} g_H , \; 
K^A_U = \gm^A_a K^a_b U^b = 0 \; ,
\end{equation}
where $g_H \equiv \alpha g$ is the renormalized surface gravity at the horizon, and
\begin{equation}
\gm_A ^c \del _c n^b \to \alpha ^{-1} \gm_A^c \del _c l^b
\Rightarrow K_A^B = \gm_A^a K_a^b \gm b^B = \alpha ^{-1} k_A^B \; ,
\end{equation}
where $k_{AB}$ is the extrinsic curvature of a spacelike 2-section of 
the true horizon,
\begin{equation}
k_{AB} \equiv \gm_A^d l_{B \| d} = \f {1} {2} \pounds _{l^a} \gm _{AB}
\; ,
\end{equation}
where $\pounds _{l^a}$ is the Lie derivative in the direction of $l^a$. We 
can decompose $k_{AB}$ into a traceless part and a trace, $k_{AB} = 
\sigma_{AB} + \f {1} {2} \gm_{AB} \theta$, where $\sigma_{AB}$ is the shear
and $\theta$ the expansion of the world lines of nearby horizon surface
elements. Then
\begin{equation}
t_s^{AB} = \f {1} {8 \pi} \lf - \sigma ^{AB} + \gm^{AB} \lf \f {1} {2}
\theta + g \rt \rt \; .	\label{stress}
\end{equation}
But this is just the equation for the stress of a two-dimensional viscous
Newtonian fluid \cite{LL} with pressure $p = \f {g} {8 \pi}$, shear viscosity 
$\eta = \f {1} {16 \pi}$, and bulk viscosity $\zeta = - \f {1} {16 \pi}$.
Hence we may identify the horizon with a two-dimensional dynamical fluid, or
membrane. Note that, unlike ordinary fluids, 
the membrane has negative bulk viscosity. This would ordinarily
indicate an instability against generic perturbations triggering 
expansion or contraction. It can be regarded as reflecting
a null hypersurface's natural tendency to expand or contract
\cite{td82}. Below we shall show how for the horizon this particular 
instability is replaced with a different kind of instability.

Inserting the $A$-momentum density ${t_s}^b_a \gm ^a_A U_b = 
t^{\; U}_{s \; \; A} \equiv \pi _A$ 
into the Einstein equations, eqns. \ref{gcod}, we arrive at 
the Navier-Stokes equation,
\begin{equation}
\pounds _{\tau} \pi _A = - \del _A p + \zeta \del _A \theta + {\rm 2} \eta
\sigma ^B _{A \; \; \| B} - T^n_A \; ,
\end{equation}
where $\pounds _{\tau} \pi _A =  \f{\pl \pi _A}{\pl \tau}$ is the 
Lie derivative (which is the general-relativistic equivalent of the convective 
derivative) with respect to proper time, and
$- T^n_A = - \gm ^a_A T^c_a n_c$ is the flux of 
$A$-momentum into the black hole.

Inserting the $U$-momentum (energy) density $t^{\; a}_{s \; \; b} U_a U^b 
\equiv \Sigma = - \f {\theta}{8 \pi}$ gives
\begin{equation}
{\pounds} _{\tau} \Sigma + \theta \Sigma = - p \theta + \zeta \theta ^{\rm 2}
+ {\rm 2} \eta \sigma _{AB} \sigma ^{AB} + T^a_b n_a U^b \; , \label{energy}
\end{equation}
which is the focusing equation for a null geodesic congruence \cite{focus}. 
We might now suspect that if the analogy with fluids extends to 
thermodynamics then eqn. \ref{energy}, as the equation of energy 
conservation, must be the heat transfer equation \cite{LL} for 
a two-dimensional 
fluid. Writing the expansion of the fluid in terms of the area, $\Delta A$, of
a patch,
\begin{equation}
\theta = \f {1} {\Delta A} \f {d \Delta A} {d \tau} \; ,
\end{equation}
we see that we can indeed re-write eqn. \ref{energy} as the heat 
transfer equation (albeit with an additional relativistic term on the left),
\begin{equation}
T \lf \f {d \Delta S} {d \tau} - \f {1} {g} \f {d^2 \Delta S} {d \tau ^2} \rt =
\lf \zeta \theta ^2 + 2 \eta \sigma _{AB} \sigma ^{AB} + T^a_b n_a U^b \rt 
\Delta A \; ,	\label{heat}
\end{equation}
with $T$ the temperature and $S$ the entropy, provided that the 
entropy is given by
\begin{equation}
S = \eta \f {k_B} {\hbar} A \; , 	\label{ent}
\end{equation}
and the temperature by
\begin{equation}
T = \f {\hbar} {8 \pi k_B \eta} g \; ,	\label{temp}
\end{equation}
where $\eta$ is some proportionality constant. 

Thus, the identification of the horizon
with a fluid membrane can be extended to the thermodynamic domain.
Nonetheless, the membrane is an unusual fluid.
The focusing equation itself, eqn. \ref{energy}, is 
identical in form to the
equation of energy conservation for a fluid. However, because the
energy density, $\Sigma$, is proportional to the expansion, $\Sigma
= - \f{\theta}{8 \pi}$, one obtains a non-linear first-order
differential equation for $\theta$ which has no counterpart for ordinary
fluids. The crucial point is that, owing to the black hole's
gravitational self-attraction, the energy density is negative, and the
solution to the differential equation represents a horizon that 
grows with time.
For example, the source-free solution with a time-slicing for which the
horizon has constant surface gravity is
\begin{equation}
\theta \lf t \rt = \f {2 g} {1 + \lf \f{2 g}{\theta \lf t_0 \rt} - 1 \rt 
e^{g \lf t_0 - t \rt }} \; .
\end{equation}
Because of the sign of the exponent, this would represent an ever-expanding
horizon if
$\theta \lf t_0 \rt$ were an initial condition; the area of the
horizon, which is related to $\theta$ by $\theta = \f{d}{d \tau} \ln
\sqrt{\gamma} \,$, expands exponentially with time. To avoid this runaway, 
one must impose ``teleological boundary conditons'' (that is, final 
conditions) rather than initial conditions. 
Hence, the horizon's growth is actually
acausal; the membrane expands to intercept infalling matter that is yet
to fall in \cite{mp}. 
This is because the membrane inherits the global character 
of the true horizon: the stretched horizon covers
the true horizon whose location can only be determined by tracking null
rays into the infinite future. 
In fact, the left-hand side of the heat transfer equation,
eqn. \ref{heat}, is of the same form as that of an electron subject
to radiation reaction; the acausality of the horizon is therefore
analogous to the pre-acceleration of the electron.

At this
classical level, using only the equations of motion, the parameter 
$\eta$ in eqn. \ref{ent}
is undetermined. However, since we have an action, we hope to do
better, since the normalization in the path integral is now fixed.
By means of a Euclidean path integral, we should actually be able to
derive the Bekenstein-Hawking entropy, including the coefficient $\eta$, 
from the membrane action. We do this in a later section.

\subsection{The Axidilaton Membrane}
Another advantage of the action formulation is that it is easily 
generalized to arbitrary fields. For example, we can extend
the membrane paradigm to include the basic fields of quantum gravity.
Here we use the tree-level effective action obtained from
string theory after compactification to four macroscopic dimensions.
This action is a generalization of the
classical Einstein-Hilbert-Maxwell action to which it reduces when
the axidilaton, $\lm$ (sometimes written as $S$), is set to $\f {i}{16 \pi}$.
The action is
\begin{equation}
S[ \lm , \cc \lm, A_a, g_{ab} ] = \int d^4 x \rg \lf \f{R}{16 \pi} 
{}-  \f { | \pl \lm |^2}{2 \l2 ^2} 
+  \f {i}{4} \lf\lm F_+ ^2 - \cc \lm F_- ^2 \rt \rt \; ,
\end{equation}
where $R$ is the four-dimensional Ricci curvature scalar,
$F_{\pm} \equiv F \pm i \tilde F$ are the self- and anti-self-dual 
electromagnetic field strengths, and
$\lm \equiv \lm _1 + i \l2 = a + i e^{-2 \phi}$ is the axidilaton, with
$a$ the axion and $\phi$ the dilaton. Solutions to
the equations of motion arising from this action include electrically-
(Reissner-Nordstr\"{o}m) and magnetically-charged black holes \cite{gm,ghs}, as
well as their duality-rotated cousins, dyonic black holes \cite{fw},
which carry both electric and magnetic charge.

The equations of motion are

\begin{equation}
\del _a \lf \f {\pl ^a \lm}{\l2 ^2} \rt + i \f {| \pl \lm |^2}{\l2 ^3} 
{} - \f {i}{2} F_- ^2 = 0 \; ,
\end{equation}
and
\begin{equation}
\del _a \lf \lm F_+^{ab} - \cc \lm F_-^{ab} \rt = 0 \; ,
\end{equation}
besides the Einstein equations.

As before, we require the external action to vanish on its own. Integration
by parts on the axidilaton kinetic term leads to a variation at the boundary,
\begin{equation}
\int d^3 x \rh \lf \d \lm \lf \f {n_a \pl ^a \cc \lm} {2 \l2 ^2} \rt + \d \cc 
\lm \lf \f {n_a \pl ^a \lm} {2 \l2 ^2} \rt \rt \; ,
\end{equation}
where $n^a$ is again chosen to be outward-pointing.
To cancel this, we add the surface term
\begin{equation}
S_{surf} = \int d^3 x \rh \lf \lm \cc q + \cc \lm q \rt \; ,
\end{equation}
so that
\begin{equation}
q = - \f {n_a \pl ^a \lm} {\l2 ^2} \; . \label{q}
\end {equation}

To interpret this, we note that the kinetic term in $\lm$ is invariant under 
global $SL \lf 2, I\!\!R \rt$ transformations of the form
\begin{equation}
\lm \to \f {a \lm + b}{c \lm + d} \; , \; \; ad - bc = 1 \; ,
\end{equation}
which are generated by Peccei-Quinn shifts, $\lm_1 \to \lm_1 + b$,
and duality transformations, $\lm \to - \f{1}{\lm}$. The Peccei-Quinn
shift of the axion can be promoted to a classical local symmetry to
yield a N\"other current:
\begin{equation}
J^a_{P-Q} = - \f{1}{2 \lm_2^2} \lf \pl ^a \lm + \pl ^a \cc \lm \rt \; .
\end{equation}
Therefore, under a Peccei-Quinn shift,
\begin{equation}
\d S_{surf} = \int d^3 x \rh \d \lm \lf q + \cc q \rt
= \int d^3 x \rh \d \lm \lf n_a J^a_{P-Q} \rt \; .
\end{equation}
The sum of the $q$ and $\cc q$ terms induced at the 
membrane, eqn. \ref{q}, is the normal component of the Peccei-Quinn current.
Hence, at the membrane,
\begin{equation}
\lf h^a_b J^b_{P-Q} \rt _{|a} = - F \tilde F - \del_a \lf \lf q + \cc q \rt 
n^a \rt \; .
\end{equation}
That is, the membrane term $\del_a \lf \lf q + \cc q \rt n^a \rt$ 
augments the dyonic $F \tilde F$ term as a source for 
the three-dimensional Peccei-Quinn
current, $h^a_b J^b_{P-Q}$, at the membrane.

The membrane is again dissipative with the Peccei-Quinn charge accounting
for the dissipation in the usual $\alpha \to 0$ limit. The local rate of
dissipation is given by the bulk stress tensor at the membrane:
\begin{equation}
T_{ab} U^a n^b = \f{1}{16 \pi} \f {\pl_a \lm \pl_b \cc \lm + \pl_a \cc \lm 
\pl_b \lm} {2 \lm_2 ^2} U^a n^b \to \f{\lm_2^2 |q|^2}{16 \pi} \; .
\end{equation}

In addition, the presence of the axidilaton affects the 
electromagnetic membrane.
(The gravitational membrane is unaffected since the surface terms come from
the Ricci scalar which has no axidilaton factor.) The electromagnetic
current is now

\begin{equation}
j_s^a = - 2 i \lf \lm F_+ ^{ab} - \cc \lm F_- ^{ab} \rt n_b \; .
\end{equation}
The surface charge is therefore
\begin{equation}
\sigma = 4 \lf \lm _2 E_\perp + \lm _1 B_\perp \rt \; ,
\end{equation}
and the surface current is
\begin{equation}
\vec{j}_s = 4 \lf \lm _2 \hat{n} \times \vec{B}_{\|} - \lm _1 \hat{n} 
\times \vec{E}_{\|} \rt \; ,
\end{equation}
which, by the regularity of the electromagnetic field, eqn. \ref{reg},
satisfies
\begin{equation} 
\lf \begin{array}{c} j_s^{\theta} \\ j_s^{\phi} \end{array} \rt =
4 \lf \begin{array}{cc} \lm _2 & \lm _1 \\ - \lm _1 & \lm _2 \end{array} \rt
\lf \begin{array}{c} E^{\theta} \\ E^{\phi} \end{array} \rt \; .
\end{equation}
The conductivity is now a tensor. When the axion is absent, the resistivity
is
\begin{equation}
\rho = \f {1} {4 \lm _2} \; .
\end{equation}
The inverse dependence on $\lm_2$ is to be expected on dimensional
grounds. The pure dilaton action can be
derived from Kaluza-Klein compactification of pure gravity in five dimensions,
where the fifth dimension is curled into a circle of radius $e^{-2 \phi} = 
\lm _2$. In five dimensions, with $c \equiv 1$, resistance (and hence
resistivity for a two-dimensional resistor such as the membrane) has 
dimensions of inverse length.
Using the regularity condition, eqn. \ref{reg}, the rate of
dissipation, for a stretched horizon defined to have uniform lapse
$\alpha$ with respect to time at infinity, $t$, is
\begin{equation}
\f{d M_{irr}}{d t} = - \int \alpha^2 \vec{S} \cdot d \vec{A} = 
\int 4 \alpha^2 \lm_2 E^2_{\|} dA  
= \int \alpha^2 \f{\lm_2}{4 |\lm|^4} \vec{j_s}^2 dA \; ,
\end{equation}
which is the Joule heating law in the presence of an axidilaton.

\section{DISSIPATION}

Given that the bulk equations of motion are manifestly symmetric under 
time-reversal,
the appearance of dissipation, as in Joule heating and fluid viscosity, might 
seem mysterious, all the more so since it has been derived from an action.

The procedure, described here, of restricting the action to some region and 
adding surface  terms on the boundary of the region cannot be applied with
impunity to any arbitrary region: a black hole is special. 
This is because the region
outside the black hole contains its own causal past; an observer who remains
outside the black hole is justified in neglecting (indeed, is unaware of)
events inside. However, even ``past sufficiency'' does not adequately
capture the requirements for our membrane approach. For 
instance, the past light cone of a spacetime point obviously
contains its own past, but an observer in this light cone must eventually
leave it. Rather, we define the notion of a future dynamically closed set:
\begin{quote}
A set $S$ in a time-orientable globally hyperbolic 
spacetime $(M, g_{ab})$ is {\em future dynamically closed} if
$J^-(S) = S$, and if, for some foliation
of Cauchy surfaces $\Sigma _t$ parameterized by the values
of some global time function, we have that $\forall \; t_0 \; \forall \; 
p \, \epsilon \, \lf S \cap \Sigma _{t_0} \rt \; \forall \; (t > t_0) \;
\exists \; q \, \epsilon \,  \lf I^+(p) \cap S \cap \Sigma _t \rt$.
\end{quote}
That is, $S$ is future dynamically closed if it contains
its own causal past and if from every point in $S$ it is possible for
an observer to remain in $S$.
Classically, the region outside the true horizon of a black hole is
dynamically closed. So too is the region on one side of a null plane in flat
space; this is just the infinite-mass limit of a black hole. 
The region outside the stretched horizon is strictly speaking {\em not} 
dynamically closed since a signal originating in the thin
region between the stretched horizon and the true horizon
can propagate out beyond the stretched horizon. However, in the limit
that the stretched horizon goes to the true horizon, $\alpha \to 0$,
this region becomes vanishingly thin so that in this limit, which is in any
case assumed throughout, we are justified in restricting the action.

The breaking of time-reversal symmetry comes from the definition of the
stretched horizon; the region exterior to the black hole does not remain
future dynamically closed under time-reversal. In other words, we have divided
spacetime into two regions whose dynamics are derived from two different
simultaneously vanishing actions, $\d \lf S_{out} + S_{surf} \rt
= \d \lf S_{in} - S_{surf} \rt = 0$. 
Given data on some suitable achronal subset we can, for the exterior 
region, predict the future but not the entire past, while, inside the
black hole, we can ``postdict'' the past but cannot determine the
entire future. Thus, our choice of the horizon as a boundary implicitly
contains the irreducible logical requirement for dissipation, that is,
asymmetry between past and future.

Besides the global properties that logically permit one to write down a
time-reversal asymmetric action, there is also a local property of
the horizon which is the proximate cause for dissipation, namely that the 
normal
to the horizon is also tangential to the horizon. Without this crucial
property - which manifests itself as the regularity condition, or
the identification of the stretched horizon
extrinsic curvature with intrinsic properties of the true horizon - there
would still be surface terms induced at the stretched horizon, but no
dissipation.

The regularity condition imposed at the boundary is not an operator
identity, but a statement about physical states: all tangential
electromagnetic fields as measured by a fiducial observer must be
ingoing. Such a statement is not rigorously true. For any given value
of $\alpha = \lf 1 - \f{2M}{r} \rt ^{\f{1}{2}}$, there is a maximum wavelength,
$\lm_{max}$, for outgoing modes that are invisible to the observer:
\begin{equation}
\lm_{max} = \f {r - 2M}{\lf 1 - \f{2M}{r} \rt ^{\f{1}{2}}} \rightarrow 2
M \alpha \; .
\end{equation}
Dissipation occurs in the membrane paradigm because the finite but
very high-frequency modes that are invisible to the fiducial observer 
are tacitly assumed not to exist. The regularity condition amounts to a
coarse-graining over these modes. It is conceivable that for
a theory with benign ultraviolet behavior, the amount of information
lost is finite. Einstein gravity is not such a theory, but one may
ask abstractly whether an effective horizon theory could exist at
a quantum level \cite{thooft,stu}. Quantum effects cause the black
hole to emit radiation. In order to preserve time-evolution unitarity,
we might require the emitted radiation to be correlated with the
interior state of the black hole. 
In this case, the membrane viewpoint remains valid only as a classical
description, since quantum-mechanically the
external universe receives information from the black hole in the form
of deviations of the radiation from thermality; the crucial premise that
the outside universe is emancipated from the internal state of the black
hole is violated. It is important to emphasize, however, that
correlations between the radiation and the horizon itself do not 
preclude the membrane paradigm. Indeed, the fact that the Bekenstein-Hawking
entropy is proportional to the surface area of the black hole suggests that,
even at the quantum level, an effective horizon theory may not be unfeasible.

\section{THE THERMODYNAMIC MEMBRANE}
To make contact with thermodynamics, we perform an analytic continuation to
imaginary time, $\tau = i t$, so that the path integral 
of the Euclideanized action becomes 
a partition function. For a stationary hole,
regularity (or the removal of a 
conical singularity) dictates a period $\beta = \int d \tau = \f{2 \pi}{g_H}$ 
in imaginary time \cite{gh}, where $g_H$ is the surface gravity; for a
Schwarzschild hole, $\beta = 8 \pi M$. 
This is the inverse Hawking temperature in
units where $\hbar = c = G = k_B = 1$. 
The partition function is then the path integral over all Euclidean metrics
which are periodic with period $\f{2 \pi}{g_H}$ in imaginary time. Since the
dominant contribution to the path integral comes from the classical solution,
we can evaluate the partition function in a stationary phase approximation:
\begin{equation}
Z = \int D g_E^{ab} \exp \lf - \f{1}{\hbar} \lf S^E_{out}[g_E^{ab}] 
+ S^E_{surf}[h_E^{ab}] \rt \rt \approx \exp \lf - \f{1}{\hbar}
\lf S^E_{out}[g^{ab}_{E \, cl}] + S^E_{surf}[h^{ab}_{E \, cl}] \rt 
\rt .
\end{equation}

The external action itself can be written as $S_{out} = S_{bulk} + S_{\infty}$,
where $S_{bulk}$ is zero for a black 
hole alone in the universe. The boundary term $S_{\infty}$ is the integral
of the extrinsic curvature of the boundary of spacetime. In fact, a term
proportional to the surface area at infinity can be included in $S_{\infty}$
without affecting the Einstein equations since the metric is held fixed at
infinity during variation. In particular, the proportionality 
constant can be chosen so that the action for all of spacetime is 
zero for Minkowski space:
\begin{equation}
S_{\infty} = \f {1}{8 \pi} \int d^3 x \rh [K] \; ,
\end{equation}
where $[K]$ is the difference in the trace of the extrinsic curvature at
the spacetime boundary for the metric $g_{ab}$ and the flat-space metric $\eta 
_{ab}$. With this choice, the path integral has a properly normalized
probabilistic interpretation. The Euclideanized value of $S_{\infty}$ for
the Schwarzschild solution is then \cite{gh}
\begin{equation}
S^E_{\infty} = \lim_{r\to\infty} \f{1}{8 \pi} \lf - 32 \pi ^2 M \rt \lf \lf 2r 
- 3M \rt -2r \lf 1 - \f{2M}{r} \rt ^{\f{1}{2}} \rt = + 4 \pi M^2 \; .
\label{Sbound}
\end{equation}

To obtain an explicit action for the membrane, we must integrate
its variation, eqn. \ref{cancel}:
\begin{equation}
\d S_{surf}[h^{ab}] = - \f {1}{16 \pi} \int d^3 x \rh \lf K h_{ab} - K_{ab}
\rt \d h^{ab} \; .
\end{equation}
We see that
\begin{equation}
S_{surf}[h^{ab}] = \int d^3 x \rh \lf B_{ab} h^{ab} - b \rt \; ,
\end{equation}
is a solution, provided that the (undifferentiated) source terms are 
$B_{ab} = + \f {1}{16 \pi} K_{ab}$ and $b = - 
\f {1}{16 \pi} K$. This action has the form of surface matter plus a
negative cosmological constant in three dimensions. 
The value of the membrane action 
for a solution to the classical field equations is then
\begin{equation}
S_{surf}[h^{ab}_{cl}] = + \f {1} {8 \pi} \int d^3 x \sqrt{-h_{cl}} 
K_{cl} 	\; . \label{Ssurf}
\end{equation}

To evaluate this, we can take our fiducial world-lines $U^a$ to be 
normal to the \mbox{isometric} time-slices of constant Schwarzschild time. 
The stretched horizon is then a surface of
constant Schwarzschild $r$. Hence $\alpha = \lf 1 -\f{2M}{r} \rt ^{\f {1}{2}}$,
$\theta = 0$, and $K = g + \theta = g$, the unrenormalized surface gravity
of the stretched horizon. Inserting these into eqn. \ref{Ssurf}, we 
obtain for the Euclidean action
\begin{equation}
S^E_{surf} = \lim_{r\to r_H} \f{1}{8 \pi} \lf \int - d \tau \rt 
\alpha 4 \pi r^2 g = - \pi {r_H}^2 = - 4 \pi M^2 \; ,
\end{equation}
where $r_H = 2M$ is the black hole's radius, and $g_H = \alpha g = \f{1}{4M}$ 
is its renormalized surface gravity.

The Euclidean membrane action exactly cancels the external action, eqn.
\ref{Sbound}. Hence the entropy is zero! 
That, however, is precisely what makes the
membrane paradigm attractive: to an external observer, 
there is no black hole - only a membrane - and so neither a 
generalized entropy nor a strictly obeyed second law of thermodynamics. 
The entropy of the
outside is simply the logarithm of the number of quantum states of the
matter outside the membrane. This number decreases as matter leaves the
external system to fall through and be dissipated by the membrane. When all
matter has fallen into the membrane, the outside is in a single state - vacuum
{}- and has zero entropy, as above.

To recover the Bekenstein-Hawking entropy, we must then use not the
combination of external and membrane actions, which gave the entropy of the
external system, but the combination of the {\em internal} and 
membrane actions,
\begin{equation}
Z_{B-H} = \int D g_E^{ab} \exp \lf - \f{1}{\hbar}\lf S^E_{in}[g_E^{ab}] -
S^E_{surf}[h_E^{ab}] \rt \rt \; ,
\end{equation}
where now $S_{surf}$ is subtracted (see eqn. \ref{split}). With $S_{in} =
\int d^4 x \rg R = 0$, the partition function for a Schwarzschild hole in
the stationary phase approximation is
\begin{equation}
Z_{B-H} \approx \exp \lf - \f{1}{\hbar} \lf + 4 \pi M^2 \rt \rt \; ,
\end{equation}
from which the Bekenstein-Hawking entropy, $S_{B-H}$, immediately follows:
\begin{equation}
S_{B-H} = \beta \lf M + \f{\ln Z_{B-H}}{\beta} \rt 
= 8 \pi M \lf M - \f{1}{8 \pi M} 4 \pi M^2 \rt = \f{1}{4} A \; ,
\end{equation}
which is the celebrated result.

For more general stationary (Kerr-Newman) holes, the Helmholtz free energy
contains additional ``chemical potential'' terms corresponding to the
other conserved quantities, $Q$ and $J$,

\begin{equation}
F = M - TS - \Phi Q - \Omega J \; ,	\label{free}
\end{equation}
where $\Phi = \f{Q}{r_+}$ and $\Omega = \f{J}{M}$, where $r_+$ is the
Boyer-Lindquist radial co-ordinate at the horizon. For a charged hole,
the action also contains electromagnetic terms. The surface electromagnetic
term, eqn. \ref{charge}, has 
the value $\f{1}{4 \pi} \int d^3 x \rh F^{ab} A_a n_b$.
However, in order to have a regular vector potential, we must gauge transform
it to $A_a = \del _a \lf t - \Phi \rt$ so that $A_a$ vanishes on the surface.
Hence, the surface action is again given by the gravitational term, which
has the Euclideanized value $S^E_{surf} = - \pi {r_+}^2$. It is easy to
verify using eqn. \ref{free} that this again leads to a black hole entropy 
equal to one-fourth of the horizon surface area, and an external entropy
of zero.

For non-stationary black holes, the extrinsic
curvature also includes a term for the expansion of the horizon, $K = g +
\theta$. Inserting this into the surface action enables us to calculate the
instantaneous entropy as matter falls into the membrane in a non-equilibrium
process. Of course, like the horizon itself, the entropy grows acausally.

\section{HAMILTONIAN FORMULATION}
The equations of motion can equally well be derived within a 
Hamiltonian formulation. This involves first singling out a global time
co-ordinate, $t$, for the external universe, which is then sliced 
into space-like surfaces, $\Sigma _t$, of constant 
$t$. We can write in the usual way
\begin{equation}
t^a \equiv \lf \f{d}{dt} \rt ^a = \alpha U^a - v^a \; ,
\end{equation}
where $U^a$ is the unit normal to $\Sigma _t$, $U^2 = -1$, and $\alpha$ and
$- v^a$ are Wheeler's lapse and shift, respectively, with $v^a = \f{dx^a}{dt}$
the ordinary 3-velocity of a particle with world-line $U^a$.
For convenience we choose the stretched horizon to be a surface of constant
lapse so that $\alpha$, which goes to zero at the true horizon, serves
as the stretched horizon regulator. The external 
Hamiltonian for electrodynamics, 
obtained from the Lagrangian via a Legendre transform and written in
ordinary three-dimensional vector notation, is
\begin{equation}
H_{out}[\phi, \vec{A}, \vec{\pi}] = \f{1}{4 \pi} \int_{\Sigma_t} 
d^3 x \sqrt{^3 g} \lf \f{1}{2} \alpha
\lf \vec{E} \cdot \vec{E} + \vec{B} \cdot \vec{B} \rt + \vec{v} \cdot
\lf \vec{E} \times \vec{B} \rt - \phi \lf \vec{\del} \cdot \vec{E} \rt
\rt \; , 	\label{ham}
\end{equation}
where $^3 g_{ab}$ is the 3-metric on $\Sigma_t$, 
$\phi \equiv -A_a t^a$ is the scalar potential, $\vec{A}_a \equiv 
{^3 g_{a}^{\; b} A_b}$
is the three-dimensional vector potential, and $\vec{\pi}^a \equiv - 
\sqrt{^3 g} \vec{E}^a$ its canonical momentum conjugate. Note that
$E^a = F^{ab} U_b$ is the co-moving electric field; $\vec{E}$ and $\vec{B}$
above refer to the fields measured by a fiducial observer with world-line
$U^a$.
Finally, the scalar potential is non-dynamical; its
presence in the Hamiltonian serves to enforce Gauss' law as a constraint.
The equations of motion are now determined by Hamilton's equations and the
constraint:
\begin{equation}
\f{\d H}{\d \vec{\pi}} = \dot{\vec{A}} \; , \; \; \f{\d H}{\d \vec{A}} 
= - \dot{\vec{\pi}} \; , \; \; \f{\d H}{\d \phi} = 0 \; .
\end{equation}
In the bulk these equations are simply Maxwell's equations but, because 
of the inner boundary, the usually discarded surface terms that arise 
during integration by parts now need to be canceled.
It is easy to show then that the above equations hold only
if additional surface terms are added to the Hamiltonian:
\begin{equation}
H = H_{out} - \int d^2 x \sqrt{\gm} \, j_s \cdot A \; .
\end{equation}
For Maxwell's equations to be satisfied in the bulk, the surface terms
are once again the surface charges and currents necessary to terminate
the normal electric and tangential magnetic fields at the stretched horizon.
Thus, the membrane paradigm is recovered.

However, it is perhaps more interesting to proceed in a slightly different
fashion. Instead of adding new terms, we can use the external Hamiltonian
to prove the validity of a principle of minimum heat production.
Such a principle, which holds under rather general circumstances for
stationary dissipative systems, holds for black holes also in
slightly non-stationary situations.

Now the time derivative of the external Hamiltonian is not zero, again
because of the inner boundary. We can use Hamilton's
equations to substitute expressions for the time derivative of the field
and its momentum conjugate. Hamilton's equations are
\begin{equation}
\dot{\vec{A}} = - \alpha \vec{E} + \vec{v} \times \vec{B} - \vec{\del} \phi
\label{adot}
\end{equation}
\begin{equation}
\dot{\vec{E}} = \vec{\del} 
\times \lf \alpha \vec{B} + \vec{v} \times \vec{E} \rt \; ,
\end{equation}
so that, making repeated use of the vector identity
\begin{equation}
\vec{\del} \cdot \lf \vec{V} \times \vec{W} \rt = \vec{W} \cdot \lf \vec{\del}
\times \vec{V} \rt - \vec{V} \cdot \lf \vec{\del} \times \vec{W} 
\rt \; ,
\end{equation}
we obtain for the energy loss
\begin{equation}
{}- \dot{H} = - \f{1}{4 \pi} \int d^2 x \sqrt{\gm} \, 
\lf \hat{n} \cdot \lf \alpha \vec{E}_{\|} \times \alpha \vec{B}_{\|} \rt
+ \vec{v} \cdot \lf E_{\perp} \alpha \vec{E}_{\|} 
+ B_{\perp} \alpha \vec{B}_{\|} \rt \rt	\; .	\label{dHdt}
\end{equation}
So far, we have used only Hamilton's equations. It
remains, however, to implement the constraint. Hence we may regard $- \dot{H}$
as a functional of the Lagrange multiplier, $\phi$. We therefore have
\begin{equation}
{}- \f{\d \dot{H}}{\d \phi} = - \f{d}{dt} \f{\d H}{\d \phi} = 0 \; .
\end{equation}
That is, the equations of motion follow from minimizing the rate of
mass increase of the black hole with respect to the scalar potential. This
is an exact statement; we now show that this reduces to a minimum heat
production principle in a quasi-stationary limit.
Now we note that the first law of black hole thermodynamics
allows us to decompose the mass change into irreducible and rotational parts:
\begin{equation}
\f{d M}{d t} = \f{d Q}{d t} + \Omega_H \f{d J}{d t} \; ,
\end{equation}
where $\Omega_H$ is the angular velocity at the horizon, and $J$ is the
hole's angular momentum. Since $|\vec{v}| \to \Omega_H$ at the horizon, we see
that the second term on the right in eqn. \ref{dHdt} corresponds to the
torquing of the black hole. When this is small, we may approximate the
mass increase as coming from the first, irreducible term. Hence, in the
quasi-stationary limit, for a slowly-rotating black hole, 
the black hole's rate of 
mass increase is given by the dissipation of external energy.
Invoking the regularity condition, eqn. \ref{reg}, then gives
\begin{equation}
D[\phi] = + \f{1}{4 \pi} \int d^2 x \sqrt{\gm} \, \lf \alpha \vec{E}_{\|} \rt 
^2 \; , \; \; \f{\d D}{\d \phi} = 0 \; ,
\end{equation}
where $\alpha \vec{E}_{\|}$ is given by eqn. \ref{adot}.
This is the principle of minimum heat production: minimizing the dissipation
functional leads to the membrane equation of motion.

We observe that we could have anticipated this answer. 
The numerical value of the Hamiltonian 
is the total energy of the system as measured at spatial infinity (assuming 
an asymptotically flat spacetime). The time derivative is then
simply the rate, as measured by the universal time of distant observers,
that energy changes. The rate of decrease of energy is the integral of the
Poynting flux as measured by local observers, multiplied by two powers of
$\alpha$, one power to convert local energy to energy-at-infinity, and one
power to convert the rate measured by local clocks to the rate measured at
infinity. Thus we can immediately define a dissipation functional:
\begin{equation}
D[\phi] \equiv - \f{1}{4 \pi} \int d^2 x \sqrt{\gm} \, \hat{n} \cdot
 \lf \vec{E}_H \times \vec{B}_H \rt \; ,
\end{equation}
where the subscript $H$ denotes that a power of $\alpha$ has been absorbed
to renormalize an otherwise divergent fiducial quantity.

In this manner, we can easily write down the dissipation functional 
for gravity for
which time-differentiating the Hamiltonian is a much more laborious exercise.
The local rate of energy transfer is given by the right-hand side of the heat
transfer equation, eqn. \ref{heat}. The
Hamiltonian for gravity satisfies two constraint equations with the lapse
and shift vector serving as Lagrange multipliers. Since the membrane 
picture continues to have a gauge freedom associated with time-slicing,
the constraint equation associated with the lapse is not implemented.
This implies that the dissipation is a functional only of the shift. Hence
we have
\begin{equation}
D[v^A] = \int d^2 x \sqrt{\gm} \lf \zeta \theta _H ^2 + 2 \eta 
\sigma _H ^2 + \alpha^2 T^a_b n_a U^b \rt \; ,
\end{equation}
where again the two powers of $\alpha$ have been absorbed to render
finite the quantities with the subscript $H$. Extremizing $D$ with respect
to $v^A$ leads to the membrane equations of motion, enforcing the 
gauge constraint or, equivalently, obeying the principle of minimum
heat production.

\section{CONCLUSION}
We have derived the equations for the membrane paradigm of black holes
from an action principle directly by demanding that both terms in
eqn. \ref{split}
are stationarized separately.  This brings advantages of conceptual unity
and ease of generalization over the traditional approach of
manipulating the equations of motion.  Specifically, the derivation
makes it clear why a membrane picture, including dissipative behavior,
is possible.  A fundamental advantage of having an action principle is
the guidance it offers for quantization--a property we used to fix the
constant in the Bekenstein-Hawking formula.

\section{APPENDIX}
In this appendix, we shall prove that eqn. \ref{zero} is zero in the limit
that the stretched horizon approaches the true horizon. In that limit, $\alpha
n^a \to l^a$. We shall make liberal use of Gauss' theorem, the
Leibniz rule, and the fact that $h^{ab} n_b = K^{ab} n_b = 0$. 
In order
to use Gauss' theorem, we note that since the ``acceleration'',
$a^c \equiv n^d \del _d n^c$ of the normal vector 
(not to be confused with the fiducial acceleration $U^d \del _d U^c$) is zero,
the 4-covariant divergence and the 3-covariant divergence of a vector in
the stretched horizon are equal, eqn. \ref{div}.

Now, variations in the metric that
are in fact merely gauge transformations can be set to zero. Using a vector
$v^a$ where $v^a$ vanishes on the stretched horizon, 
we can gauge away the variations
in the normal direction so that $\d g_{ab} \to \d h_{ab}$. Then
the left-hand side of eqn. \ref{zero} becomes
\begin{eqnarray}
\lefteqn{\int d^3 x \rh h^{bc} \lf \del _a \lf n^a \d h_{bc} \rt -
\del _c \lf n^a \d h_{ab} \rt \rt } & & \nonumber \\ 
& = & \int d^3 x \rh \lf \del _a \lf h^{bc} n^a \d h_{bc} \rt -
\lf \del _a h^{bc} \rt n^a \d h_{bc} 
{}- \del _c \lf h^{bc} n^a \d h_{ab} \rt +
\lf \del _c h^{bc} \rt n^a \d h_{ab} \rt \nonumber \\
& = &\int d^3 x \rh \lf \del _a \lf h^{bc} n^a \d h_{bc} \rt 
+ \lf n^c a^b + n^b a^c \rt \d h_{bc}
{}- \lf h^{bc} n^a \d h_{ab} \rt _{|c} \right. \nonumber \\
& & \indent \indent \indent \left. - h^{bc} n^a \d h_{ab} a_c
{}- K n^b n^a \d h_{ab} - a^b n^a \d h_{ab} \rt \nonumber \\
& & \mbox{(using $h^{bc} = g^{bc} - n^b n^c$, $K_{ab} = + h_a^c \del _c n_b$, 
and $\del_c w^c = w^c_{|c} + w^c a_c$ for $w^c \epsilon {\cal{H}}$)} 
 \nonumber \\
& = & \int d^3 x \rh \lf \del _a \lf h^{bc} n^a \d h_{bc} \rt 
- K n^b n^a \d h_{ab} \rt \nonumber \\
& & \mbox{(using Gauss' theorem, and $a^c = 0$)} \nonumber \\
& = & \int d^3 x \rh \lf \del _a \lf h^{bc} \f {\alpha} 
{\alpha} n^a \d h_{bc} \rt
- K \lf \d \lf n^b n^a h_{ab} \rt - 
n^a h_{ab} \d n^b - n^b h_{ab} \d n^a \rt \rt \nonumber \\
& \to & \int d^3 x \rh \del _a \lf h^{bc} \f {1} {\alpha} l^a \d h_{bc} \rt 
\nonumber \\
& & \mbox{(using $h_{ab} n^b = 0$, and $\alpha n^a \to l^a$)} \nonumber \\
& = & \int d^3 x \rh \lf h^{bc} \f {1} {\alpha} l^a \d h_{bc} \rt _{|a} 
\nonumber \\
& = & 0
\end{eqnarray}

\section{ACKNOWLEDGEMENTS}
M.P. is grateful to Thibault Damour, Alexandre Polyakov, and Kip Thorne for
illuminating conversations. F.W. is supported in part by DOE grant 
DE-FG02-90ER-40542.


\begin{thebibliography}{99}

\bibitem{he}
For a review see, e.g., S. W. Hawking and G. F. R. Ellis, {\em The Large-Scale
Structure of Spacetime} (Cambridge University Press, Cambridge, England, 1973).

\bibitem{ch}
D. Christodoulou, Phys. Rev. Lett. {\bf 25}, 1596 (1970).

\bibitem{chr}
D. Christodoulou and R. Ruffini, Phys. Rev. D {\bf 4}, 3552 (1971).

\bibitem{bek}
J. D. Bekenstein, Phys. Rev. D {\bf 7}, 2333 (1973).

\bibitem{swh}
S. W. Hawking, Phys. Rev. Lett. {\bf 26}, 1344 (1971).

\bibitem{hr}
R. S. Hanni and R. Ruffini, Phys. Rev. D {\bf 8}, 3259 (1973).

\bibitem{td78}
T. Damour, Phys. Rev. D {\bf 18}, 3598 (1978).

\bibitem{zna}
R. L. Znajek, Mon. Not. R. Astron. Soc. {\bf 185}, 833 (1978). Znajek's
approach differs from ours in that he considers a membrane with a
non-zero thickness and volume conductivities.

\bibitem{td79}
T. Damour, th\`{e}se de doctorat d'\'{e}tat, University of Paris VI, 1979
(unpublished).

\bibitem{td82}
T. Damour, in {\em Proceedings of the Second Marcel Grossman Meeting on
General Relativity}, edited by R. Ruffini (North-Holland, Amsterdam, 1982),
p. 587.

\bibitem{tmac}
D. A. Macdonald and K. S. Thorne, Mon. Not. R. Astron. Soc. {\bf 198},
345 (1982).

\bibitem{pt}
R. H. Price and K. S. Thorne, Phys. Rev. D {\bf 33}, 915 (1986).

\bibitem{mp}
{\em Black Holes: The Membrane Paradigm}, edited by
K. S. Thorne, R. H. Price, and D. A. Macdonald (Yale University Press,
London, 1986). It should be noted that these authors imply by ``the membrane
paradigm'' not only the existence of a stretched horizon, but also a
3+1 split of spacetime. Their motivation for severing space from time is
pragmatic: astrophysical processes are usually not modeled in full tensor
formalism. In this paper, we require such a split only to write down
Ohm's law, the Joule law, and the Navier-Stokes equation, in their familiar
Galilean form.

\bibitem{carlip}
S. Carlip, Phys. Rev. D {\bf 51}, 632 (1995).

\bibitem{tunnel}
M. K. Parikh and F. Wilczek, in preparation.

\bibitem{mtw}
C. W. Misner, K. S. Thorne, and J. A. Wheeler, {\em Gravitation} (Freeman,
New York, 1973).

\bibitem{LL}
See, e.g., L. D. Landau and E. M. Lifshitz, {\em Fluid Mechanics} (Pergamon,
Oxford, 1987), \S 49.

\bibitem{focus}
S. W. Hawking, Phys. Rev. Lett. {\bf 26}, 1344 (1971).

\bibitem{gm}
G. Gibbons and K. Maeda, Nucl. Phys. {\bf B298}, 741 (1988).

\bibitem{ghs}
D. Garfinkle, G. Horowitz, and A. Strominger, Phys. Rev. D {\bf 43}, 3140 
(1991).

\bibitem{fw}
A. Shapere, S. Trivedi, and F. Wilczek, Mod. Phys. Lett. A {\bf 6}, 2677 
(1991).

\bibitem{thooft}
G. 't Hooft, Nucl. Phys. {\bf B335}, 138 (1990).

\bibitem{stu}
L. Susskind, L. Thorlacius, and J. Uglum, Phys. Rev. D {\bf 48}, 3743 (1993).

\bibitem{gh}
G. W. Gibbons and S. W. Hawking, Phys. Rev. D {\bf 15}, 2752 (1977).

\end{thebibliography}
\end{document}